# The continuum limit in the quenched approximation*

C. Bernard,[a] T. Blum,[b] C. DeTar,[c] Steven Gottlieb,[d] Urs M. Heller,[e] J. Hetrick,[b] K. Rummukainen,[d] R. Sugar,[f] D. Toussaint,[b] and M. Wingate,[g]

[a]Department of Physics, Washington University, St. Louis, MO 63130, USA
[b]Department of Physics, University of Arizona, Tucson, AZ 85721, USA
[c]Physics Department, University of Utah, Salt Lake City, UT 84112, USA
[d]Department of Physics, Indiana University, Bloomington, IN 47405, USA
[e]SCRI, The Florida State University, Tallahassee, FL 32306-4052, USA
[f]Department of Physics, University of California, Santa Barbara, CA 93106, USA
[g]Physics Department, University of Colorado, Boulder, CO 80309, USA

Previous work at $6/g^2 = 5.7$ with quenched staggered quarks is extended with new calculations at 5.85 and 6.15 on lattices up to $32^3 \times 64$. These calculations allow a more detailed study of extrapolation in quark mass, finite volume and lattice spacing than has heretofore been possible. We discuss how closely the quenched spectrum approaches that of the real world.

## 1. INTRODUCTION

It would certainly be exciting to calculate the observed spectrum of light hadrons from first principles. Although there have been many lattice spectrum QCD calculations [1], the large nucleon to rho mass ratio has been a persistent problem which may only go away as the lattice spacing shrinks to zero. In order to make significant progress on the light quark spectrum, very high statistics are needed to understand and control the systematic errors. Three physical parameters, the volume $V$, the quark mass $am_q$ and the lattice spacing $a$, must all be adjusted in such a way that approaching the desired values is more computationally demanding.

This work reports on a series of large scale calculations with quenched staggered quarks [2]. We also summarize some results from the literature for quenched Wilson quarks and dynamical staggered quarks for purposes of comparison. Of course, the two quenched formulations should agree in the continuum limit, and the success of the quenched approximation is determined by how well they agree with experiment.

After summarizing our runs, we will compare the finite volume effects that we see at 5.7 with those at weaker coupling [3,4]. We then discuss the extrapolation in quark mass and the nucleon to rho mass ratio as a function of lattice spacing [5]. The mass ratio is extrapolated to zero lattice spacing and compared with the observed value.

Because we have studied a wide range of couplings and quark masses, we are able to examine the evidence for quenched chiral logarithms [6] in the pion mass [7]. In addition, we have recently begun to study the spectrum with an alternative gluon action [8] and summarize our results there by showing an Edinburgh plot.

## 2. PARAMETERS OF CALCULATION

The results presented at Lattice '94 have been extended quite significantly. For most of the runs the number of lattices has doubled. In addition to increasing the maximum spatial size studied at $6/g^2 = 5.7$ from $N_s = 20$ to 24, we have done extensive calculations at $6/g^2 = 5.85$ and 6.15. The weakest coupling run is not completed. Results shown here supersede what was shown at the conference, but should still be considered preliminary. For each run we used five quark masses. For 5.7 and 5.85, we used $am_q = 0.16, 0.08, 0.04, 0.02$ and 0.01. For 6.15, the masses were halved. On each lattice we calculated hadron propagators from wall sources separated by eight time

*presented by S. Gottlieb

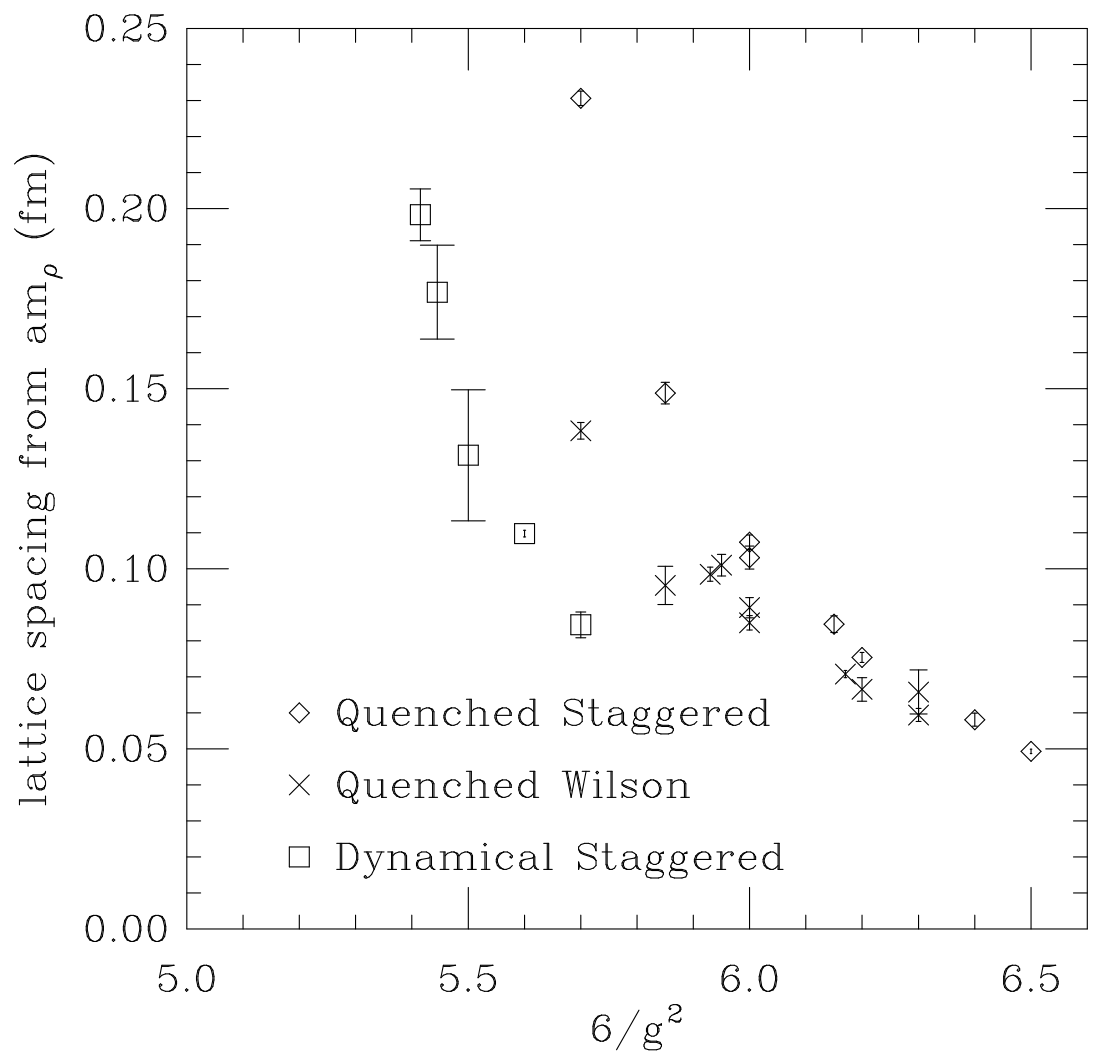

Figure 1. Lattice spacing from the rho mass as a function of gauge coupling.

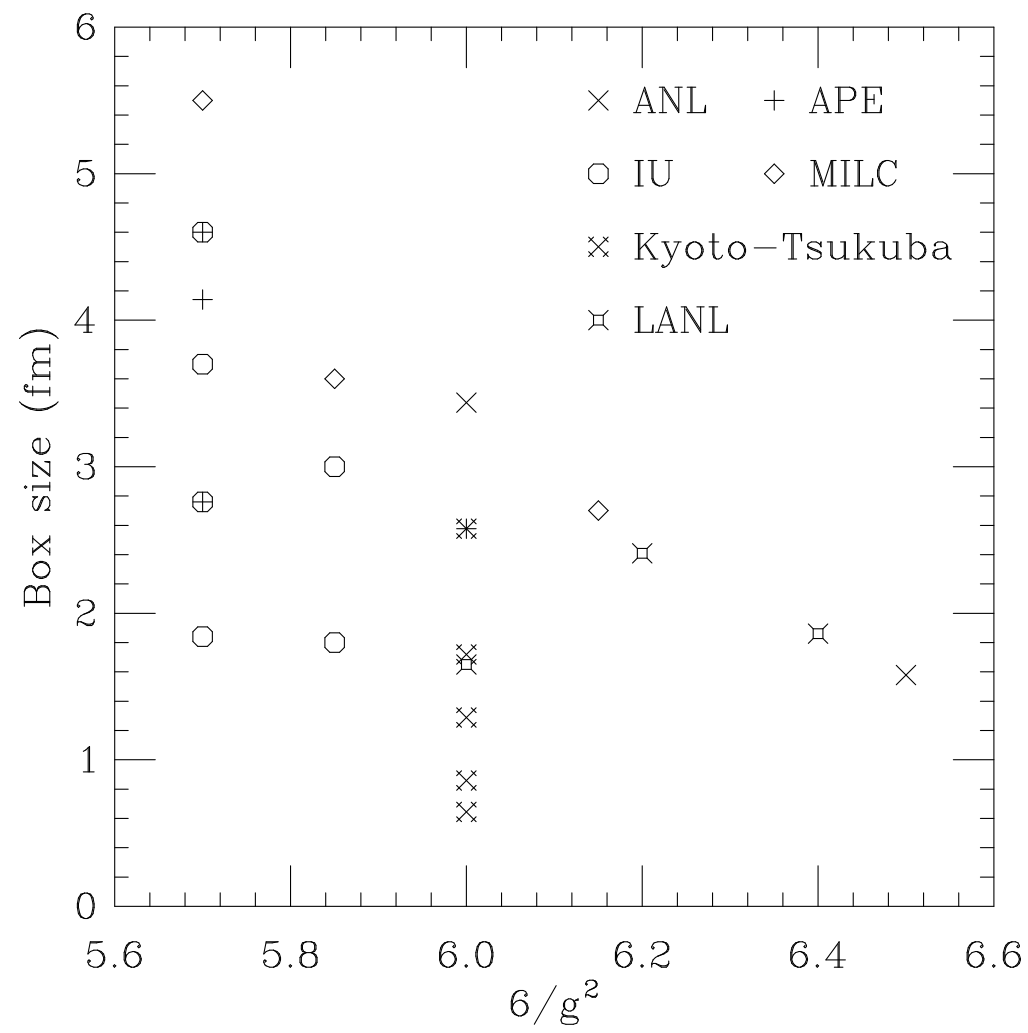

Figure 2. Box size in fm of quenched staggered calculations.

slices. For the two stronger couplings, we have $N_t = 48$ and six sets of hadron propagators per lattice. For the weak coupling, $N_t = 64$ and we have eight propagators per lattice.

Table 1
Number of lattices analyzed

| $6/g^2 = 5.7$ | $N_s^3 \times 48$ | |
|---|---|---|
| $N_s$ | Lat '94 | Lat '95 |
| 8 | 400 | 600 |
| 12 | 205 | 400 |
| 16 | 205 | 400 |
| 20 | 90 | 200 |
| 24 | | 200 |
| $6/g^2 = 5.85$ | $N_s^3 \times 48$ | |
| 12 | | 200 |
| 20 | | 200 |
| 24 | | 200 |
| $6/g^2 = 6.15$ | $N_s^3 \times 64$ | |
| 32 | | 105 |

## 3. FINITE VOLUME EFFECTS

When everything is expressed in terms of lattice units, it is easy to lose track of the physics. We use the $\rho$ mass for zero quark mass to set the the lattice spacing for each coupling. In Fig. 1 we show the lattice spacing $a$ as a function of gauge coupling for quenched staggered and Wilson quarks and for dynamical staggered quarks. We note that the large discrepancy between the scale as determined from the Wilson and staggered quarks goes away as $a \to 0$. We also note that the range of lattice spacing studied with the two types of quarks is not all that different. The quenched calculations with $6/g^2 \geq 6.4$ probably all suffer from finite size effects. With dynamical fermions this is true in most cases for the calculations with $6/g^2 \geq 5.6$.

Knowing the lattice spacing, we can now look at the physical box size used in each calculation. Restricting our attention to quenched staggered calculations [2–4,9–12] this is shown in Fig. 2. For $6/g^2 = 5.7$ and 5.85, the maximum box size in lattice units is 24. For weaker couplings, the largest box size is 32. Information on finite size effects can come only from three couplings and for 5.7 and 6.0 the most sizes have been studied.

Two approaches can be taken to finite size effects. One is to gain a thorough understanding of them at strong coupling where the calculations are easier. We expect that the effects will depend upon the quark mass, with greater effects expected for light quarks. If the effects are physical, the mass dependence should depend on $m_\pi/m_\rho$ and the effect should scale with lattice spacing. With this understanding, it should be possible to do a calculation at weaker coupling and small box size but correct for the box size. A second approach to the finite size effects is to determine what box size leads to effects that are smaller than a tolerable error and then always use boxes that size or larger.

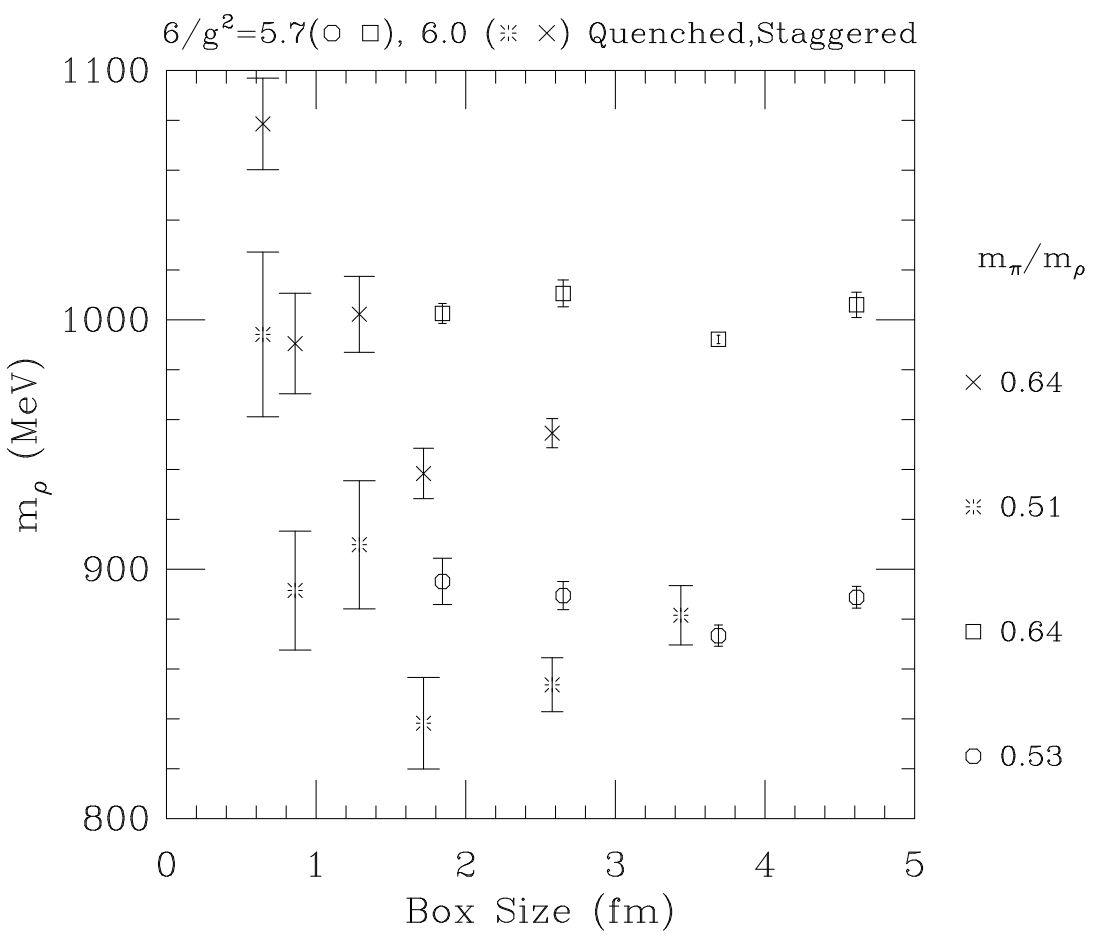

Figure 3. Rho mass *vs* box size for two couplings.

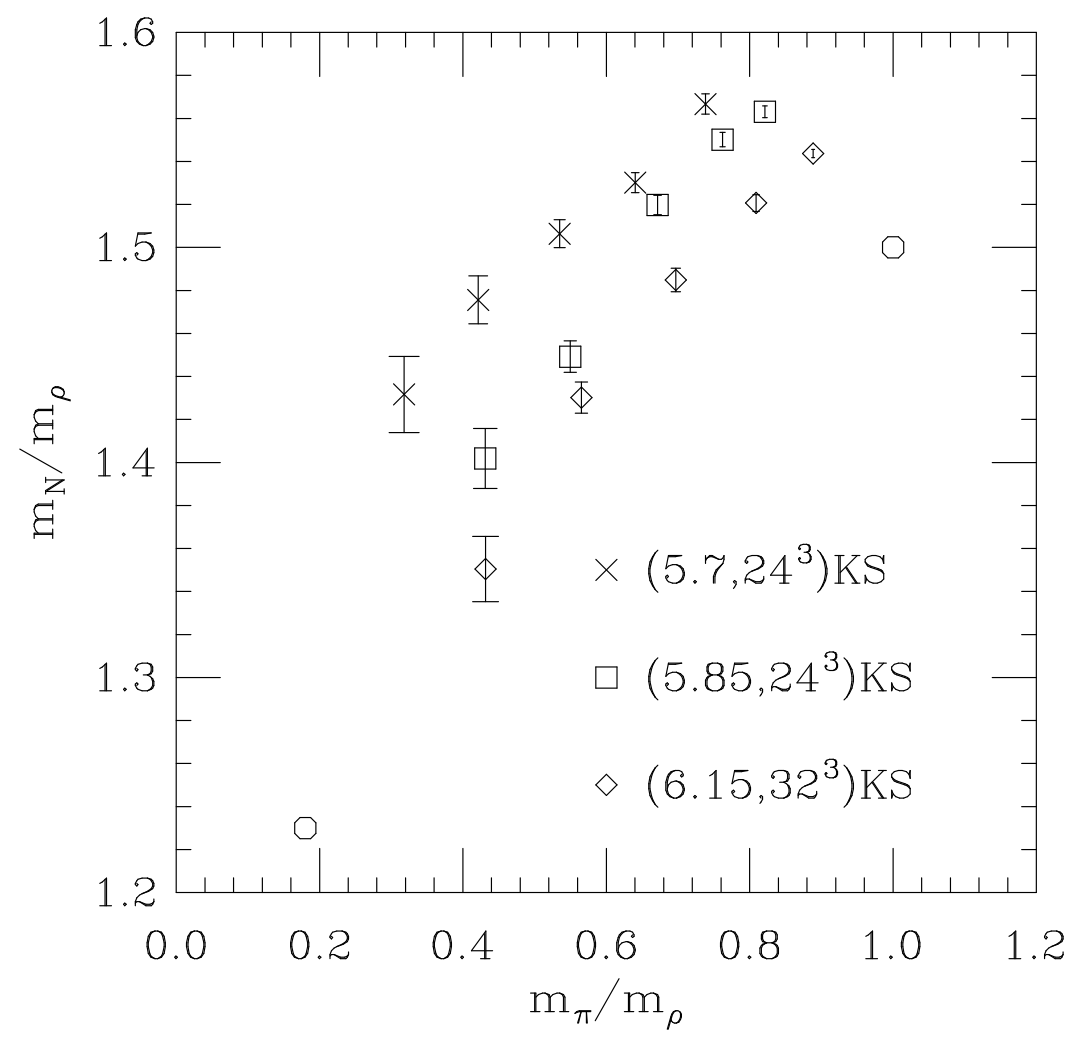

Figure 4. Edinburgh plot comparing $6/g^2 = 5.7$, 5.85 and 6.15.

Figure 3 shows what is known about the rho mass at $6/g^2 = 5.7$ and 6.0 where several box sizes have been studied. In each case, results for two quark masses are shown. The heavier quark mass corresponds to $m_\pi/m_\rho = 0.64$ and the lighter mass to 0.53 and 0.51 for the stronger and weaker couplings, respectively. On the basis of what is currently known, it does not appear that the first approach can yet be applied. Smaller errors are required at 6.0 and smaller volumes at 5.7 before a better comparison can be made. Smaller errors may soon be available at 6.0[7].

We have also looked at finite size effects on the nucleon and for smaller quark masses. We conclude that for quenched staggered quarks any box size below 2 fm is getting uncomfortable. For quenched Wilson quarks there is less evidence to rely upon as no more than three volumes have been studied for any particular hopping parameter and coupling. Previous work with dynamical quarks favors a box size $\geq 2.5$ fm for that case.

## 4. EXTRAPOLATION IN MASS, $a$

Because all our calculations are done with unphysically large quark masses, it is necessary to extrapolate toward the chiral limit. However, before we do so we may look at the Edinburgh plot and see that the nucleon to rho mass ratio is clearly dropping as we increase the coupling from 5.7 to 6.15. (See Fig. 4.)

We have tried several fits for the quark mass dependence of the hadron masses. We have a wide range of masses, and find that linear fits to the $\rho$ and nucleon masses for all five quark masses are not good fits. We have also attempted quadratic fits to all five masses and linear fits to the lightest three or four. As an example, for $6/g^2 = 5.85$ and $N_s = 24$, we find $m_\rho(m_q = 0) = 0.588(3)$ based on a quadratic fit. ($\chi^2 = 1.5$ for 2 dof.) Using a linear fit for the three lightest masses, we find 0.593(4) with $\chi^2 = 2.6$ for 1 dof. We are having great difficulty getting good fits for the nucleons and pions. For the parameters detailed above, $m_N(m_q = 0) = 0.806(4)$ and $\chi^2 = 35$ for 2 dof, based on a quadratic fit. A linear fit to the lighest three masses yields $m_N(m_q=0)=0.800(5)$ and $\chi^2=6$ for 1 dof. (The fits we have done do not yet include the full covariance matrix for the different quark masses, so perhaps one should be unhappy with these $\chi^2$ values even for the $\rho$.) Our fits do not take into account the possibility of quenched chiral logarithms. We discuss evidence for such effects in the pion mass below. We are actively investigating how to improve the reliability of the mass extrapolation.

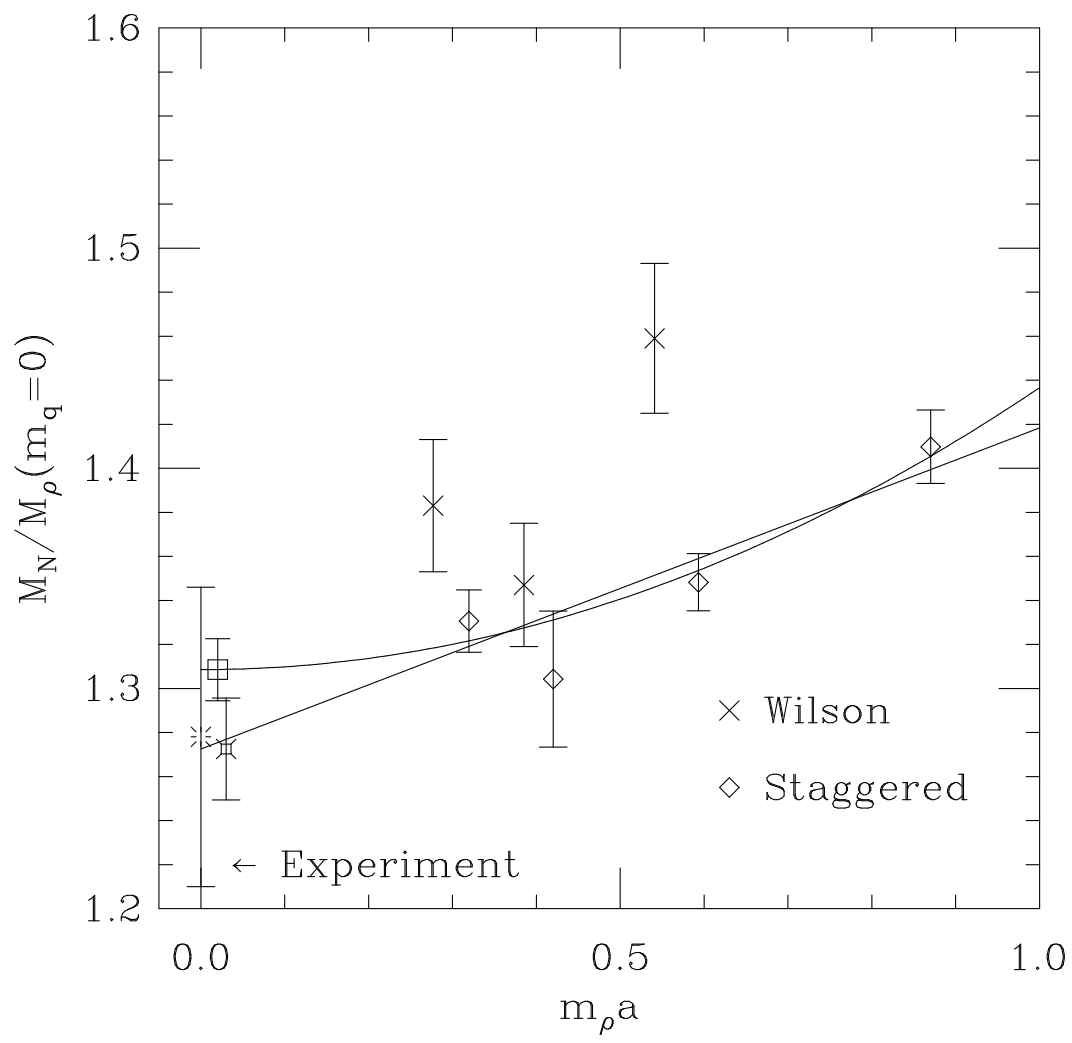

Figure 5. Extrapolation of chiral limit of nucleon to rho mass ratio.

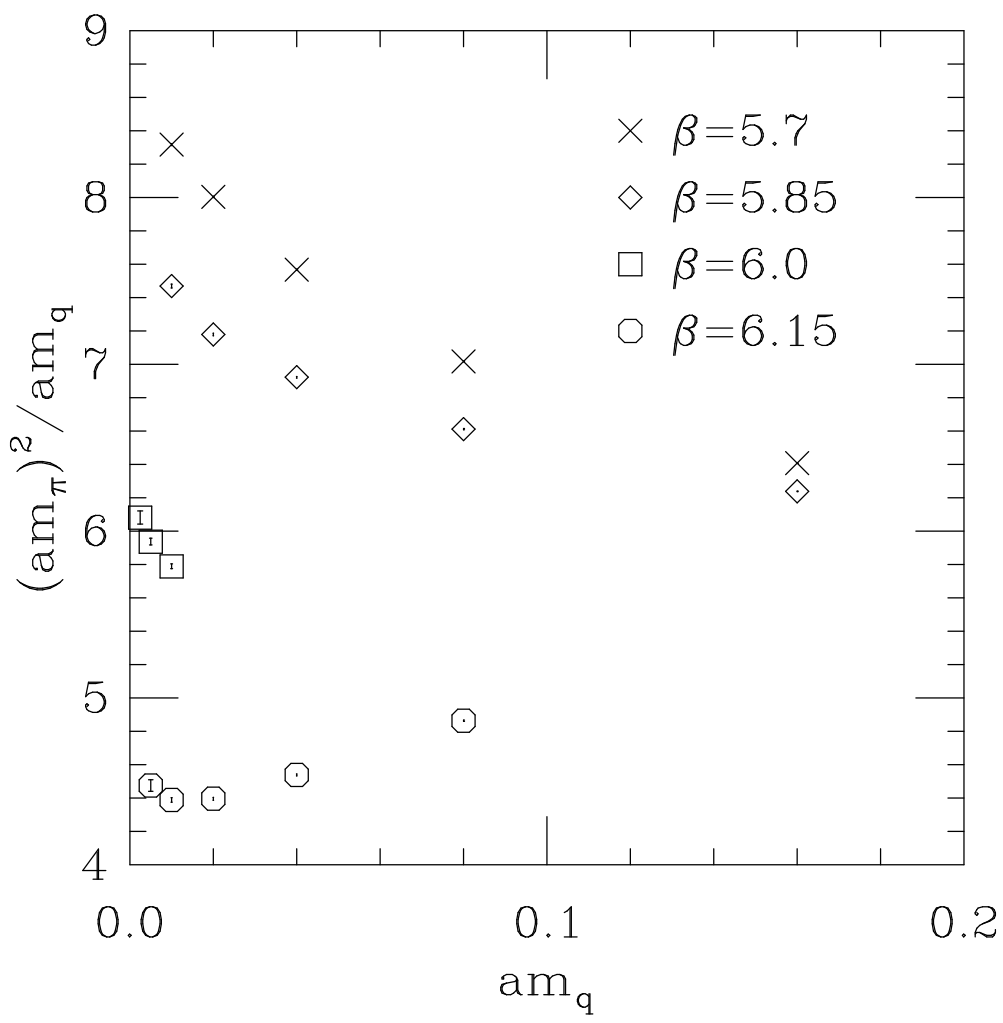

Figure 6. $m_\pi^2/m_q$ in lattice units.

With the nucleon and rho masses extrapolated to zero quark mass, we can now plot the ratio of these two masses as a function of lattice spacing. In Fig. 5, we plot this ratio *vs* the lattice rho mass and compare with Wilson quarks [5]. The points from $6/g^2=5.7$, 5.85 and 6.15 are taken from our own calculations. For 6.0, we have taken masses from Refs. [3,4] and extrapolated to zero quark mass. We show two extrapolations to zero lattice spacing for the staggered calculation. A linear fit and a constant plus quadratic are shown. The kinetic energy operator for the staggered quarks is supposed to have corrections of order $a^2$. Thus we expect the quadratic fit is relevant. Plotted as a burst is the extrapolated value for the Wilson quarks. The three extrapolated values are slightly displaced from $am_\rho=0$ so that the errors can easily be seen; however, all extrapolations are to 0. These values should be compared with the experimental value of 1.22.

The nucleon to rho mass ratio for quenched staggered quarks now appears to be in at least as good shape as it is for Wilson quarks. The error should be reduced further when our run at 6.15 is complete. Other groups should soon have new results at other couplings.

## 5. QUENCHED CHIRAL LOGS

Recently, Kim and Sinclair [7] have examined the pion mass at $6/g^2=6.0$ for light quark masses 0.01 0.005 and 0.0025. They find that $m_\pi^2/m_q$ decreases as $m_q$ is increased and claim this is evidence for quenched chiral logarithms. We have studied three values of the coupling and have five quark masses for each coupling. We do not go as close to the chiral limit as they do; however, we have looked at much heavier masses. In Fig. 6 we show our results plus those of Kim and Sinclair at 6.0. We see that $m_\pi^2/m_q$ is not constant for any of the couplings studied. The falloff at 5.7 and 5.85 persists to very large $m_q$. At 6.15, the ratio is now increasing for large $m_q$, however, it is decreasing for our two smaller quark masses. We merely wish to urge some caution in interpreting what is cur-rently observed as evidence for quenched chiral logarithms. We plan to study this issue further in the coming months. The reader might also find the talk by R. Mawhinney [13] to be of interest.

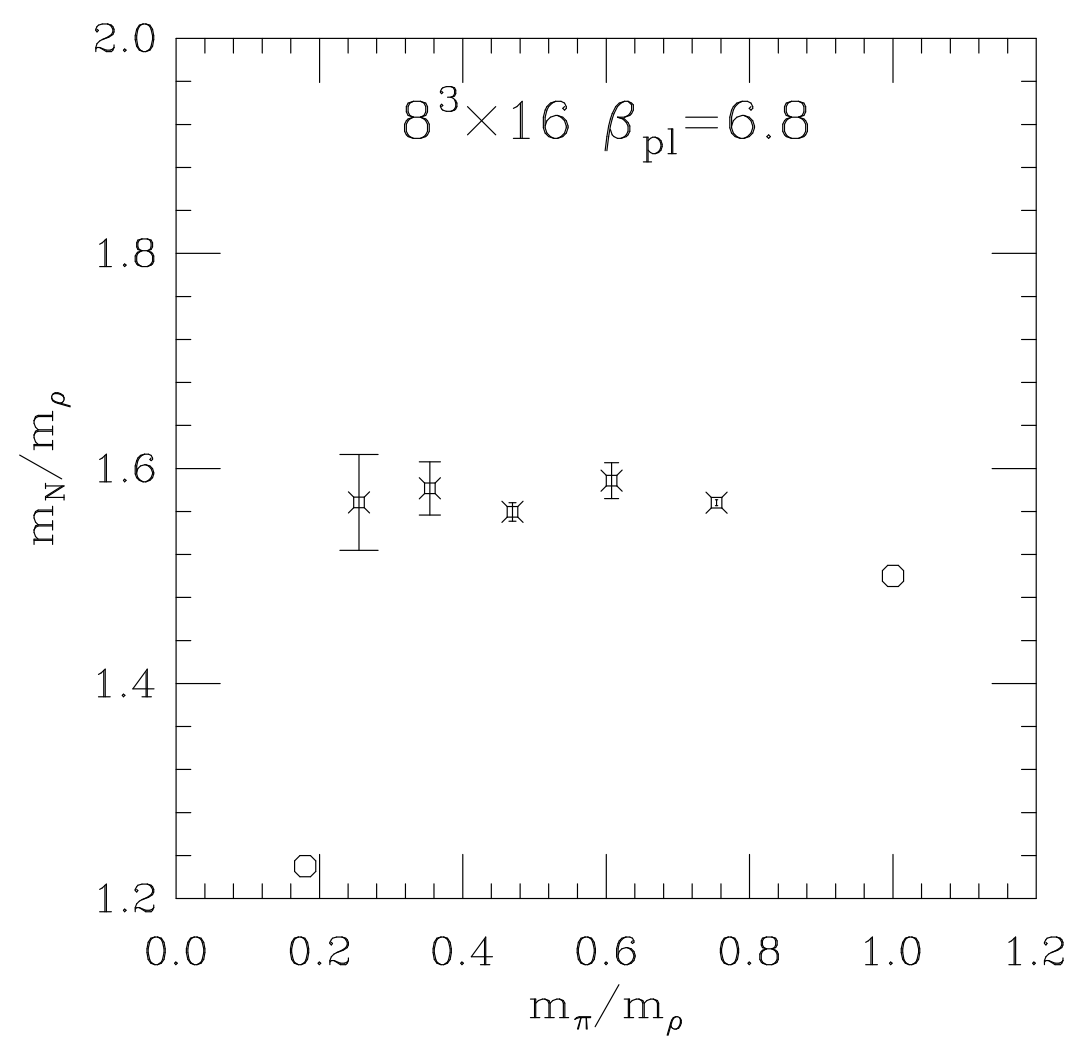

Figure 7. Edinburgh plot with an alternative action.

## 6. ALTERNATIVE ACTIONS

Very recently we have begun to study the light quark hadron spectrum on gauge fields generated with an improved action. More details of the improvement scheme can be found in the review by Lepage [8]. Using a three term action with a plaquette $\beta = 6.8$, which has been seen to correspond to a lattice spacing of 0.4 fm, we found the spectrum with Kogut-Susskind quarks. An Edinburgh plot is shown in Fig. 7. We do not find this plot markedly improved from the usual Wilson gauge action, but we caution that the important step of improving the staggered quark action has not yet been taken. For a study of the improvement with Wilson quarks with a clover improved action, see the talk by R. Edwards[14].

This work was supported by the U.S. Department of Energy and the National Science Foundation. Calculations were done on Intel Paragons at Indiana University and Sandia National Laboratory, and Cray T3Ds at Pittsburgh Supercomputer Center and NERSC. Archival storage has been provided at the National Center for Supercomputer Applications. We thank M. Golterman, P. Mackenzie, J. Smit and D. Weingarten for discussions of the $a$ dependence of the approach to the continuum.